\documentclass{aa501}

\usepackage{graphicx}
\usepackage{natbib}
\bibpunct{(}{)}{;}{a}{}{,} 

\newcommand{\La}{\mbox{${\rm Ly\alpha}$}}
\newcommand{\Line}[3]{\Ion{#1}{#2}\,$\lambda$\,#3}
\newcommand{\Lines}[3]{\Ion{#1}{#2}\,$\lambda\lambda$\,#3}
\newcommand{\Ion}[2]{#1{\,\scriptsize #2}}
\newcommand{\Mwd}{\mbox{$M_{\rm wd}$}}
\newcommand{\Teff}{\mbox{$T_{\rm eff}$}}
\newcommand{\Porb}{\mbox{$P_{\rm orb}$}}
\newcommand{\es}{\mbox{$\rm erg\;s^{-1}$}}
\newcommand{\Msun}{\mbox{$M_{\odot}$}}
\newcommand{\kms}{\mbox{$\rm km\,s^{-1}$}}
\begin{document}

\title{HST/STIS spectroscopy of the exposed white dwarf in the
short-period dwarf nova EK\,TrA
\thanks{Based on observations made with the NASA/ESA Hubble Space
Telescope, obtained at the Space Telescope Science Institute, which is
operated by the Association of Universities for Research in Astronomy,
Inc., under NASA contract NAS 5-26555.}}

\author{B.T. G\"ansicke\inst{1} \and
        P. Szkody\inst{2} \and
        E.M. Sion\inst{3} \and
	D.W. Hoard\inst{4} \and
        S. Howell\inst{5} \and
        F.H. Cheng\inst{3} \and
	I. Hubeny\inst{6}
        }
\offprints{B. G\"ansicke, e-mail: boris@uni-sw.gwdg.de}

\institute{
  Universit\"ats-Sternwarte, Geismarlandstr. 11, 37083 G\"ottingen, 
  Germany
\and
  Department of Astronomy, Box 351580, University of Washington,
  Seattle, WA 98195, USA
\and
  Department of Astronomy \& Astrophysics,  Villanova 
  University, Villanova, PA 19085, USA,
\and
  Cerro Tololo Inter-American Observatory, Casilla 603, La Serena, Chile,
\and
  Astrophysics Group, Planetary Science Institute, 620 North 6th
  Avenue, Tucson, AZ 85705.
\and
  Laboratory for Astronomy and Solar Physics, NASA/GSFC, Greenbelt, MD 20711, USA
}

\date{Received \underline{\hskip2cm} ; accepted \underline{\hskip2cm} }

\authorrunning{G\"ansicke et al.}

\abstract{
We present high resolution Hubble Space Telescope ultraviolet
spectroscopy of the dwarf nova EK\,TrA obtained in deep
quiescence. The Space Telescope Imaging Spectrograph data reveal the
broad \La\ absorption profile typical of a moderately cool white
dwarf, overlayed by numerous broad emission lines of He, C, N, and Si
and by a number of narrow absorption lines, mainly of \Ion{C}{I} and
\Ion{Si}{II}. Assuming a white dwarf mass in the range
$0.3-1.4$\,\Msun\, we derive $\Teff=17\,500-23\,400$\,K for the
primary in EK\,TrA; $\Teff=18\,800$\,K for a canonical mass of
0.6\,\Msun. From the narrow photospheric absorption lines, we measure
the white dwarf rotational velocity, $v\sin i=200\pm100$\,\kms.  Even
though the strong contamination of the photospheric white dwarf
absorption spectrum by the emission lines prevents a detailed
quantitative analysis of the chemical abundances of the atmosphere,
the available data suggest slightly sub-solar abundances.  The high
time resolution of the STIS data allows us to associate the observed
ultraviolet flickering with the emission lines, possibly originating
in a hot optically thin corona above the cold accretion disk.
}

\maketitle

\section{Introduction}
Photospheric emission from the accreting white dwarfs in non-magnetic
cataclysmic variables (CVs) was unmistakably identified for the first
time in the quiescent ultraviolet spectra of the two dwarf novae U\,Gem
and VW\,Hyi \citep{panek+holm84-1,mateo+szkody84-1}, obtained with the
\textit{International Ultraviolet Explorer} (IUE). 
While relatively good estimates for the effective temperatures of
these stars could be derived from the IUE data, a full analysis of the
properties of accreting white dwarfs in CVs --~such as photospheric
abundances, mass and rotation rate~-- had to await the availability of
high resolution and high signal-to-noise ratio ultraviolet
spectroscopy. A few CVs have been well studied with the Hubble Space
Telescope throughout the first decade of its operation, demonstrating
what overwhelming amount of information can be drawn from such
high-quality observations, and leading to important and sometimes
surprising results for our understanding of the physics of white
dwarfs, accretion disks, and novae
(e.g. \citealt{horneetal94-1,sionetal94-1,sionetal96-1,sionetal97-1},
for recent reviews see \citealt{gaensicke00-1} and
\citealt{sion99-1}).

However, the small number of systems studied with HST at a sufficient
level of detail has thus far prevented any serious statistical
analysis of the white dwarf properties in CVs and the subsequent
derivation of any correlations between the observed white dwarf
properties and global characteristics of these interacting binary
stars, such as their orbital period or the mass transfer rate. We have
initiated a large HST program, covering a total of 16 non-magnetic
CVs, to significantly enlarge the observational data base, with the
ultimate aim of accomplishing the first systematic study of the
properties of accreting white dwarfs in CVs.

In this paper, we report the analysis of the first observations
obtained in this program, unveiling the accreting white dwarf in the
SU\,UMa type dwarf nova EK\,TrA. 

\section{Observations}
\subsection{Spectroscopy}
A single HST/STIS ultraviolet spectrum of EK\,TrA was obtained on 1999
July 25, UT 18:31, using the E140M echelle grating and the
$0.2\arcsec\times0.2\arcsec$ aperture, covering the range
1125$-$1710\,\AA\ with a nominal resolution of $R\sim90000$. The
choice of the echelle grating was motivated by our goal to measure the
rotational velocity of the white dwarf in EK\,TrA, since
low-resolution G140L spectroscopy can provide only a lower limit of
$\sim300$\,\kms. The exposure time was 4302\,s, covering $\sim80$\,\%
of the 90.5\,min binary orbit. Figure\,1 shows the STIS spectrum of
EK\,TrA sampled in 0.6\,\AA\ bins.

The last outburst of EK\,TrA occured on 1999 February 20, 155 days
before the HST observations. We can, therefore, assume that the STIS
spectrum shows the system deep in quiescence. The continuum flux is a
factor $\sim3.5$ lower than in the faintest recorded IUE spectrum,
obtained during the late decline from a super-outburst
\citep{hassall85-1,gaensickeetal97-1}.

\begin{figure}
\includegraphics[angle=270,width=8.8cm]{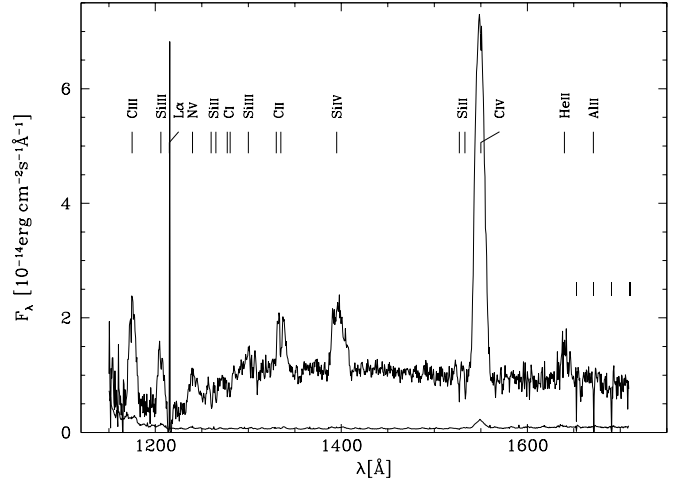}
\caption[]{\label{f-stis_spectrum}
The HST/STIS spectrum of EK\,TrA in 0.6\,\AA\ bins. The line at the
bottom gives the the error of the observation, which  varies
over each individual echelle order. The four tickmarks above the
spectrum at $\lambda>1650$\,\AA\ show the position of the
0.4--1.0\,\AA\ wide gaps between the reddest echelle orders.}
\end{figure}

Clearly visible in Fig.\,\ref{f-stis_spectrum} is the broad flux turn
over at $\lambda\la1320$\,\AA\, which we identify as the \La\
absorption line from the white dwarf photosphere. The signal-to-noise
ratio rises from $\sim3$ at the short wavelength end to a maximum of
$\sim15$ around 1400\,\AA\, and decreases to $\sim10$ at the long
wavelength end of the spectrum. In addition to the broad \La\
absorption the spectrum contains a large number of medium- and high
excitation emission lines, i.e. \Ion{He}{II}, \Ion{C}{II,\,III},
\Ion{N}{V}, and \Ion{Si}{III,IV} (Table\,\ref{t-line_id}). The narrow
emission of \La\ is of geocoronal origin. Finally, a number of narrow
($\mathrm{FWHM}\approx1.5-2.5$\,\AA) absorption lines from
low-ionisation transitions are observed, i.e. \Ion{Si}{II} and
\Ion{C}{I} (Table\,\ref{t-line_id}). At the red end of the spectrum,
three gaps between the echelle orders are apparent at 1653\,\AA,
1671\,\AA, and 1690\,\AA.

\begin{table}
\newlength{\width}
\newcommand{\FN}[1]{\settowidth{\width}{$^{#1)}$}\mbox{$^{#1)}$}\hspace*{-\width}}
\newcommand{\fn}[1]{\mbox{$^{#1)}$}}
\begin{flushleft}
\caption{\label{t-line_id}
Line identifications in the STIS spectrum of EK\,TrA. e\,=\,emission
line, p\,=\,photospheric absorption line, i\,=\,interstellar
absorption line, :\,=\,uncertain detection.}
\begin{tabular}{rrl@{\hspace*{3ex}}rrl}\hline\noalign{\smallskip}
Ion & $\lambda$\,[\AA] & Note &
Ion & $\lambda$\,[\AA] & Note \\
\noalign{\smallskip}\hline\noalign{\smallskip}
\Ion{C}{III}   & 1176\FN{*}       & e        & \Ion{Si}{II}  & 1309\FN{*}             & p       \\
\Ion{Si}{III}  & 1207\FN{*}       & e        & \Ion{C}{II}   & 1324\FN{*}             & p       \\
\Ion{N}{V}     & 1240\FN{*}       & e        & \Ion{C}{II}   & 1335\FN{*}             & e, p, i \\
\Ion{Si}{II}   & 1260             & e:, p, i & \Ion{Si}{IV}  & 1400\FN{*}             & e       \\
\Ion{Si}{II}   & 1265\FN{*}       & e:, p    & \Ion{Si}{II}  & 1527                   & e, p, i \\
\Ion{C}{I}     & 1277\FN{*}       & p        & \Ion{Si}{II}  & 1533                   & e, p    \\
\Ion{C}{I}     & 1280\FN{*}       & p        & \Ion{C}{IV}   & 1550\FN{*}             & e       \\
\Ion{Si}{III}  & 1294--1303\FN{*} & e        & \Ion{He}{II}  & 1640\FN{*}             & e       \\
\Ion{O}{I}     & 1302             & i        & \Ion{Al}{II}  & 1671\FN{+}             & p       \\
\Ion{Si}{II}   & 1305\FN{*}       & p,i      \\
\noalign{\smallskip}\hline\noalign{\smallskip}
\multicolumn{6}{p{8cm}}{\fn{*} unresolved doublet/multiplet; \fn{+} falls near one of the 
gaps between the echelle orders.}
\end{tabular}
\end{flushleft}
\end{table}

\begin{figure}
\includegraphics[angle=270,width=8.8cm,]{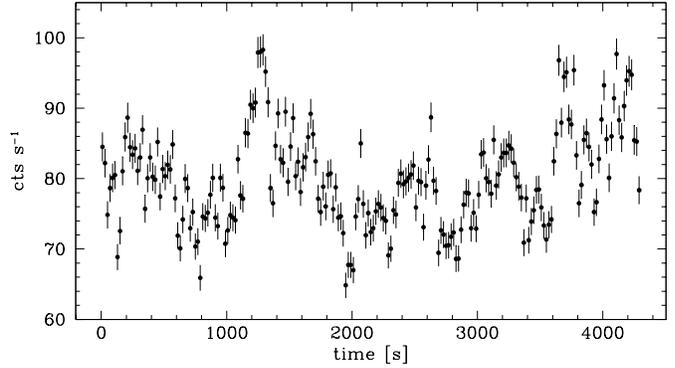}
\caption[]{\label{f-stis_lc}
STIS G140L count rate in 20\,s bins}
\end{figure}

\begin{figure*}
\includegraphics[width=4.4cm,angle=270]{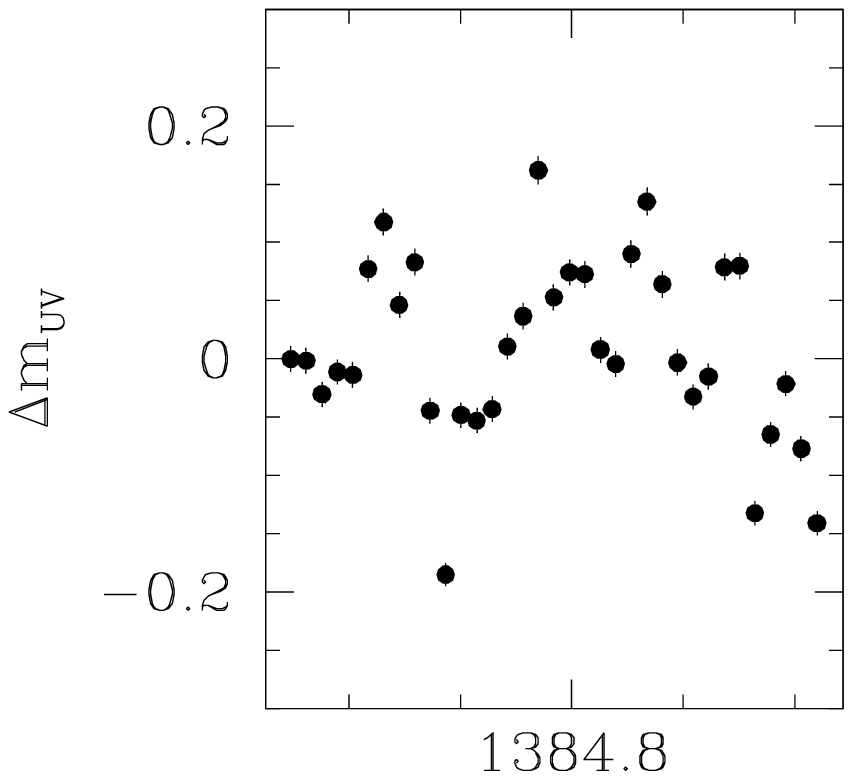}
\includegraphics[width=4.4cm,angle=270]{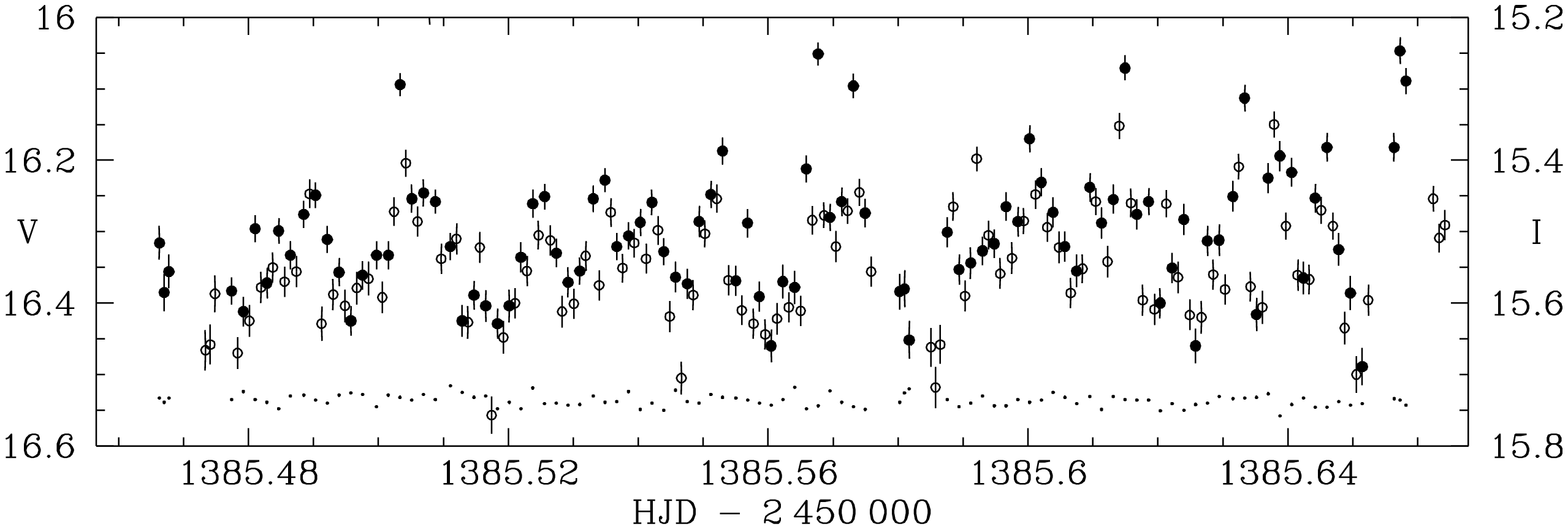}
\caption[]{\label{f-photometry}
Quasi-simultaneous ultraviolet (left), optical, and near-infrared
photometry (right, filled symbols: $V$, open symbols: $I$) of
EK\,TrA. The STIS data have been binned in 120\,s. The light curve of
the comparison star used in the reduction of the ground-based data is
shown as small dots (shifted down by 1.9\,mag).}
\end{figure*}

\subsection{\label{s-photometry}Photometry}
The STIS data were taken in the time-tagged mode, which provides
information on the temporal variability of the ultraviolet flux with a
$125\mu$\,s resolution. We have extracted the arrival time for all
photons, excluding only a small region of the detector containing the
geocoronal \La\ line. Figure\,\ref{f-stis_lc} shows the resulting
light curve binned to a time resolution of 20\,s.

We obtained ground-based photometry of EK TrA contemporaneous with the
$HST$ observations using the 0.9-m telescope at Cerro Tololo
Inter-American Observatory (CTIO).  These observations were obtained
on the night of 1999 July 25--26 UT from 23:00 to 04:00.  The data
consist of an $\approx 4.25$ hr long light curve of EK TrA, composed
of alternating 45 s exposures in $V$ and $I$
(Fig.\,\ref{f-photometry}). In addition, we obtained observations of
Landolt standard stars \citep{landolt92-1} and three sets of
individual $BVRI$ observations (each composed of three images in each
filter) of EK TrA at the start, middle, and end of the light curve.
The images were reduced in the standard fashion with IRAF tasks
utilizing zero and sky flat field images obtained on the same night.
We measured instrumental magnitudes using the IRAF task {\em qphot}
with a 10 pixel (4\arcsec) radius aperture ($\approx4\times$ the
seeing FWHM), and then calibrated them using the standard star data.
The mean values and $1\sigma$ uncertainties of the nine $BVRI$
observations obtained for EK TrA are: $B=16.261\pm0.121$,
$V=16.267\pm0.135$, $R=15.854\pm0.076$, and $I=15.590\pm0.071$.  The
uncertainties of these mean magnitudes include both random (detection
noise and flickering) and systematic (calibration) effects.  Typical
single measurement systematic uncertainties are $\sigma_{B}=0.035$
mag, $\sigma_{V}=0.019$ mag, $\sigma_{R}=0.014$ mag, and
$\sigma_{I}=0.014$ mag; typical detection noise uncertainties for the
$V$ and $I$ light curves are $\sigma_{V} = \sigma_{I} \la 0.02$ mag.

\begin{figure}
\includegraphics[angle=270,width=8.8cm]{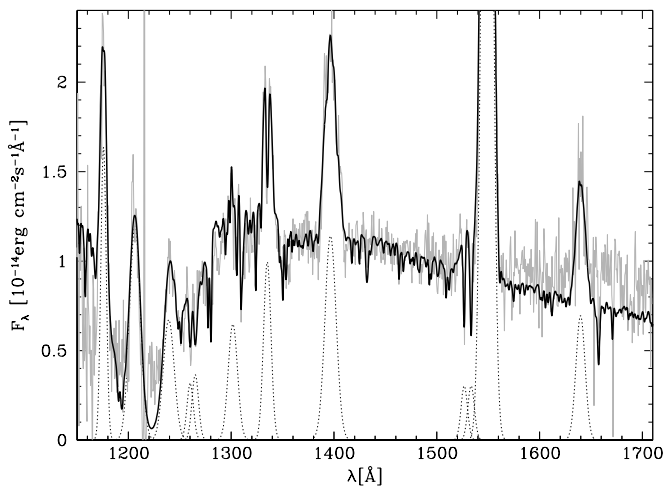}
\caption[]{\label{f-stis_fit}
The HST/STIS spectrum of EK\,TrA (gray line) along with a
($\Teff\,=\,18\,800$\,K, $\log g=8.0$) white dwarf + Gaussian emission
line model. The contribution of the Gaussian emission lines is shown
for clarity also by dotted lines.}
\end{figure}

\section{Analysis}

\cite{gaensickeetal97-1} analysed an IUE spectrum of EK\,TrA obtained
during the late decline from a superoutburst with a composite
accretion disk plus white dwarf model. They found that a white dwarf
with $\Teff\approx16\,000-20\,000$ contributes $\sim25$\,\% to the
ultraviolet flux. Considering that the first HST observation of
EK\,TrA was obtained a long time after its last outburst, and that the
ultraviolet flux corresponds quite well to the flux level of the white
dwarf predicted by G\"ansicke et al., we modelled the observed
ultraviolet spectrum with a set of white dwarf model spectra, and
neglect the possible continuum contribution of the accretion disk. The
likely contribution of the disk is discussed below.

\subsection{White dwarf effective temperature and surface gravity}
We have computed a grid of solar abundance model spectra covering
$\Teff=16\,000-20\,000$\,K in 200\,K steps and $\log g=7.25-8.50$ in
0.25 steps for the analysis of the photospheric white dwarf
emission. This spectral library was generated with the codes TLUSTY195
and SYNSPEC45 \citep{hubeny88-1,hubeny+lanz95-1}.  We fitted the model
spectra to the STIS data, allowing for Gaussian emission of
\Ion{He}{II}, \Ion{C}{II,III}, \Ion{N}{V}, and \Ion{Si}{II,III} (see
Table\,\ref{t-line_id}), and we excluded from the fit a 20\,\AA\ broad
region centered on the geocoronal \La\ emission. In order to achieve a
physically meaningful fit, we had to constrain the components of the
\Line{Si}{II}{1260,65} doublet to have the same FWHM, and the
components of \Line{Si}{II}{1527,33} to have the same FWHM and flux. 

It is, in principle, possible to derive from such a fit both the
effective temperature and the surface gravity of the white dwarf, as
both parameters determine the detailed shape of the photospheric \La\
absorption profile. Unfortunately, in our observation of EK\,TrA the
\La\ profile is strongly contaminated by various emission lines. In
addition, the pressure-sensitive $H_2^+$ transition at 1400\,\AA,
which is formed in a white dwarf photosphere with
$\Teff\la20\,000$\,K, is totally covered up by emission of
\Line{Si}{IV}{1394,1403}. The best-fit parameters $(\Teff,\log g)$ are
approximately linearly correlated, $\Teff\approx2360\times\log g-95$
with insignificant variations in $\chi^2$. As a result, it is not
possible to derive an estimate for the surface gravity, and, hence,
for the mass of the white dwarf. For the range of possible white dwarf
masses, $0.3-1.4$\,\Msun, the fit to the STIS data constrains the
white dwarf temperature to $\Teff\approx17\,500-23\,400$\,K, which
confirms the results obtained by \citet{gaensickeetal97-1}. The
scaling factor of the model spectrum provides an estimate of the
distance to EK\,TrA which depends, however, on the assumed white dwarf
mass. For a typical 0.6\,\Msun\ white dwarf, $d=200$\,pc, for the
extreme limits $\Mwd=0.3$ (1.4)\,\Msun, $d=300$ (34)\,pc.
Considering that \citet{gaensickeetal97-1} derived a lower
limit on the distance of $\sim180$\,pc from the non-detection of the
donor star --~in agreement with the 200\,pc estimated by
\cite{warner87-1} from the disk brightness~-- a massive white dwarf
($\Mwd\ga1.0\Msun$) can probably be excluded.

\begin{figure*}
\includegraphics[height=10.cm]{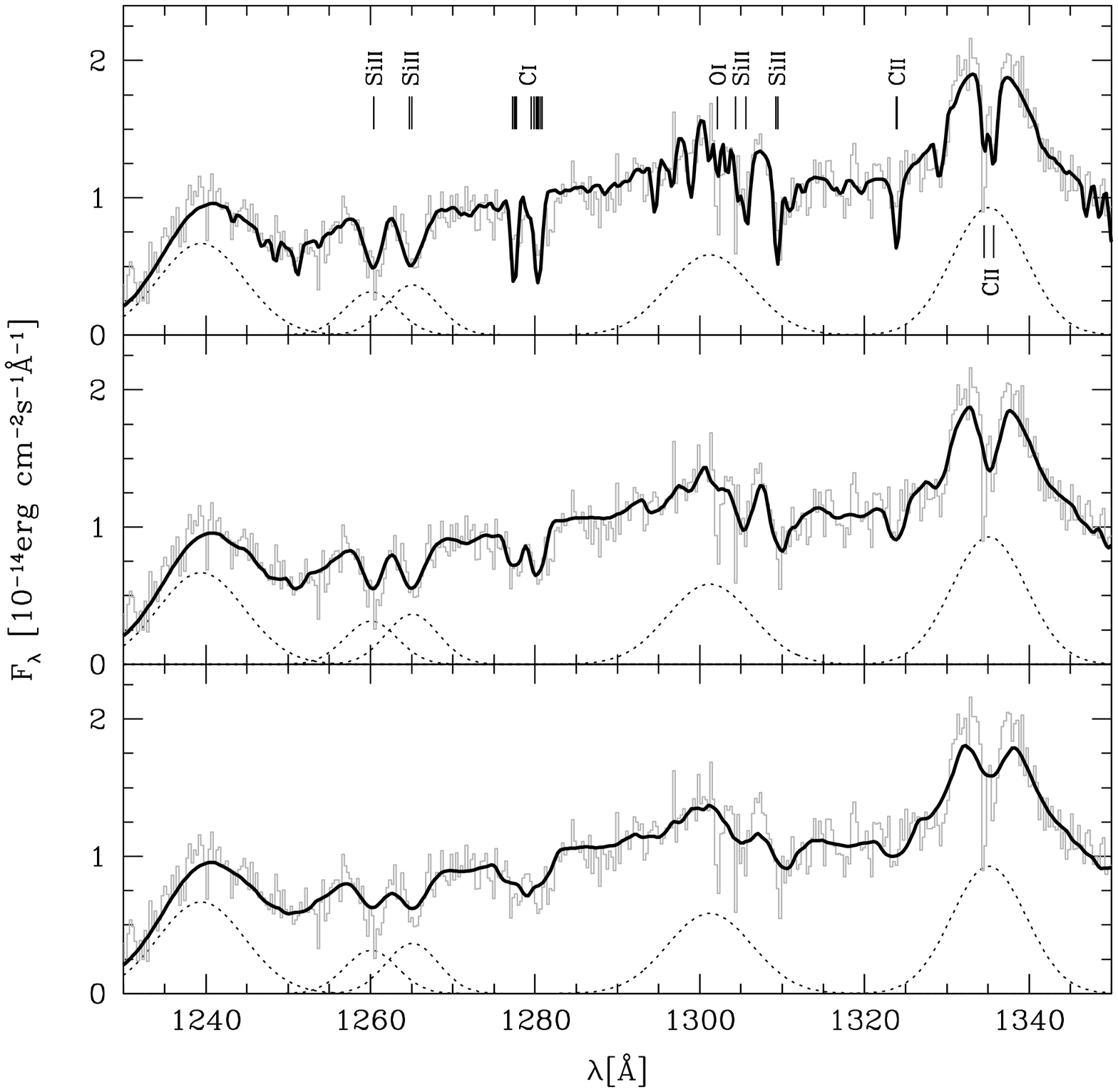}
\includegraphics[height=10.cm]{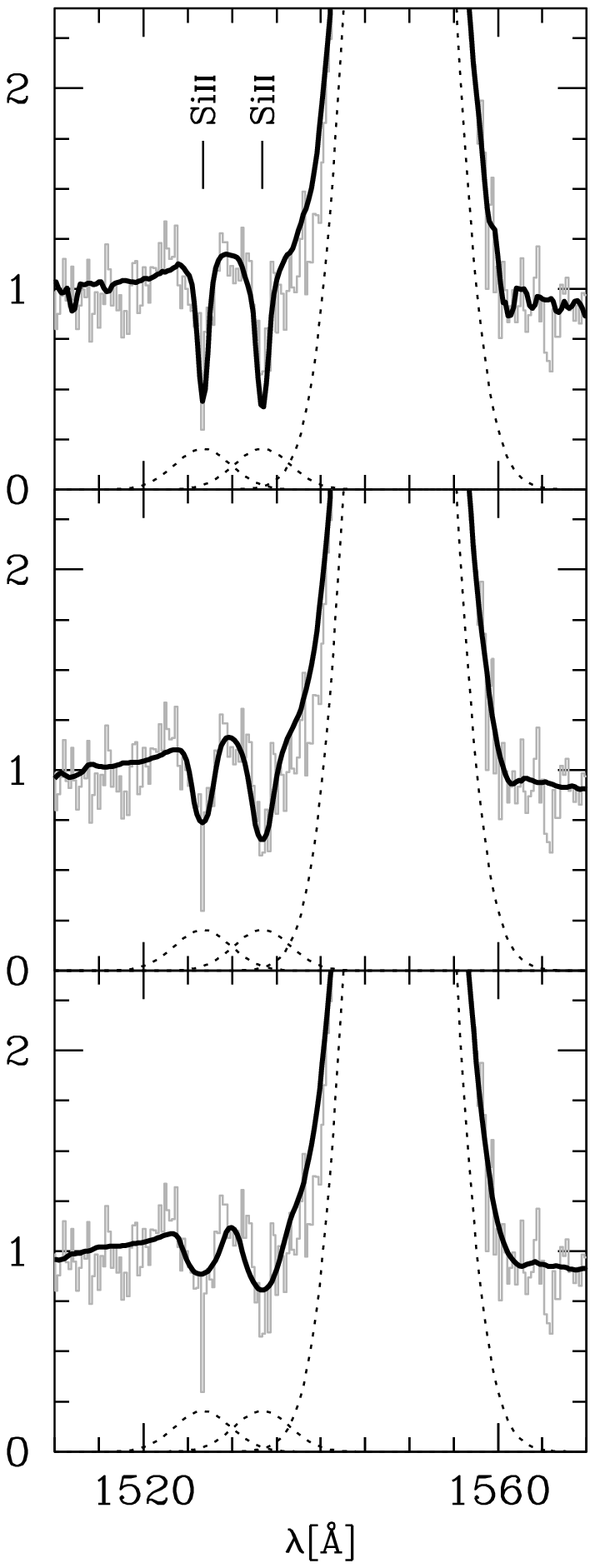}
\caption[]{\label{f-rotation}
Two regions of the STIS spectrum which contain narrow metal absorption
lines (see Table\,\ref{t-line_id}).  The STIS data and the model
spectra are sampled in 0.3\,\AA\ steps, which resolves well the
observed photospheric absorption lines.  Plotted as thick lines is a
($\Teff=18\,800$\,K, $\log g=8.0$, abundances\,=\,0.5$\times$solar)
white dwarf model, broadened for rotational velocities of 100, 300,
500\,\kms\ (from top to bottom). Note the narrow interstellar
absorption features at \Line{Si}{II}{1260}, \Line{O}{I}{1302},
\Line{Si}{II}{1304}, \Lines{C}{II}{1334,35}, and \Line{Si}{II}{1527}.}
\end{figure*}

A model fit with ($\Teff\,=\,18\,800$\,K, $\log g=8.0$), corresponding
to a white dwarf mass of $\sim0.6$\,\Msun, is shown in
Fig.\,\ref{f-stis_fit}. The observed flux exceeds the model flux by
$\sim10-15$\,\% at the red end of the STIS spectrum
($\lambda\ga1550$\,\AA).  In the optical, the model shown in
Fig.\,\ref{f-stis_fit} has $V=17.4$, which is well below the
quasi-simultaneous optical magnitude of our CTIO photometry
(Sect\,\ref{s-photometry}).  The optical flux excess over the white
dwarf contribution has been modelled by \citet{gaensickeetal97-1} with
emission from an optically thin accretion disk, providing a quite good
fit to the Balmer emission lines. It is very likely that the flux
excess at ultraviolet wavelengths is related to the same
source. Indeed, the flux contribution of an optically thin disk is
dominated in the near-ultraviolet by bound-free and its emission
increases towards longer wavelengths, reaching a maximum at the Balmer
jump.  In view of the short available wavelength range where the disk
noticeably contributes ($\sim100$\,\AA), we refrained from a
quantitative analysis of the disk contribution. Nevertheless, as a
conservative estimate, the contribution of the accretion disk to the
continuum flux at wavelengths shorter than 1550\,\AA\ is certainly
much lower than 10\%.


\subsection{Photospheric abundances and rotation rate}
A number of narrow photospheric absorption lines are observed that
clearly have an origin in the white dwarf photosphere\footnote{The
observed width of these lines corresponds to velocities of a few
hundred \kms. If these lines had an origin in the accretion disk, they
would form in regions that are hot enough to populate \Ion{Si}{II} --
i.e. in the inner disk --, where the Keplerian velocities are higher.}
(Table\,\ref{t-line_id}).  Unfortunately most of these absorption
lines are contaminated by optically thin radiation from the accretion
disk. For instance, the STIS spectrum reveals weak emission of
\Lines{Si}{II}{1527,33}. Consequently, even though no obvious
\Lines{Si}{II}{1260,65} emission is observed, we can not assume that
the photospheric absorption lines of this \Ion{Si}{II} resonance
doublet are uncontaminated. It is therefore clear that a quantitative
analysis of the photospheric absorption lines will be prone to
systematic uncertainties.

We computed a small grid of model spectra for ($\Teff=18\,800$\,K,
$\log g=8.0$) covering white dwarf rotation rates of 100--500\,\kms\
in steps of 100\,\kms\, and abundances of $1.0$, $0.5$, and $0.1$
times the solar values. These spectra were fitted to the strongest
\Ion{Si}{II} and \Ion{C}{I} lines observed in the STIS spectrum of
EK\,TrA, again allowing for Gaussian emission lines of \Ion{Si}{II}
(Table\,1). No emission lines were added for the \Ion{C}{I} features,
as \Ion{C}{I} was --~to our knowledge~-- never observed in emission in
the ultraviolet spectra of CVs.  The quality of the fits varies
somewhat depending on the considered lines, but generally indicates
$v\sin i=200\pm100$\,\kms, and sub-solar abundances. The best-fit
rotation rate does not significantly depend on the assumed
abundances. Figure\,\ref{f-rotation} shows the fits for
$0.50\times$\,solar abundances and for 100, 300, and 500\,\kms.

The very narrow cores observed in \Line{Si}{II}{1260} and
\Line{Si}{II}{1527} are of interstellar nature. Note that the
1265\,\AA\ and 1533\,\AA\ members of these two doublets do not show
such narrow cores: they are transitions from excited levels, which
are not populated in the interstellar medium. Additional strong
interstellar features found in the STIS spectrum are
\Line{O}{I}{1302}, \Line{Si}{II}{1304}, and
\Lines{C}{II}{1334,35}. All these interstellar lines are clearly
visibly also in the unbinned data. Their measured positions
agree within $\pm10$\,\kms\ with their rest wavelengths, which nicely
demonstrates the quality of the absolute wavelength calibration. The
positions of the photospheric white dwarf lines constrains the
systemic velocity of EK\,TrA to $\gamma\la70$\,\kms.

\subsection{Short-term variability}
Both the ultraviolet flux (Fig.\,\ref{f-stis_lc}) and the optical flux
(Fig.\,\ref{f-photometry}) show short-term fluctuations on time scales
much shorter than the orbital period of 90.5\,min
(15.9$\mathrm{d}^{-1}$).  We have computed Lomb-Scargle periodograms
for the all three wavelength ranges, and find significant power at
$32\textrm{d}^{-1}$, $67\textrm{d}^{-1}$, $93\textrm{d}^{-1}$ in the
ultraviolet and at $45\textrm{d}^{-1}$ and $65\textrm{d}^{-1}$ in the
optical/IR.  The white dwarf rotation rate derived above, $v\sin
i\approx200$\,\kms, implies a spin period of a few minutes, depending
somewhat on the white dwarf radius and the inclination of the system
($\sim60^{\circ}$, \citealt{mennickent+arenas98-1}). The detected
periods are much longer, and most likely represent the preferred
quasiperiodic time scale for flickering during the observations. The
fact that the $V$ and the $I$ band light curves are practically
identical both in shape and in amplitude is somewhat surprising, as
this suggests a flat spectrum for the flickering. 

In order to identify the spectral origin of the ultraviolet flickering
we constructed light curves from the STIS photon event file for the
strongest emission line, \Line{C}{IV}{1550} (using the range
1530--1565\,\AA), and for an adjacent emission line-free region (using
the range 1414--1530\,\AA). The (continuum-subtracted) \Ion{C}{IV}
light curve shows a similar variation as the total light curve
(Fig.\,\ref{f-stis_lc}), but with a much larger amplitude. For
comparison, the standard deviation from the mean count rate is
$\sigma\approx30$\,\% in the \Ion{C}{IV} light curve
vs. $\sigma\approx9$\,\% in the total light curve. The variation in
the continuum count rate extracted from the line-free region is
$\sigma\approx6$\,\%, which is comparable to the errors due to photon
statistics. This confirms the assumption that the continuum is mainly
made up of photospheric emission from the white dwarf.  Considering
that $\sim25$\,\% of the total ultraviolet flux observed with STIS is
contained in the various emission lines, we conclude that the
flickering is primarily associated with an optically thin region,
possibly some kind of corona above a cold accretion disk.
For a discussion of the possible excitation mechanisms causing
the line emission, see \citet{maucheetal97-1}. Assuming a distance of
180\,pc, the luminosity of the ultraviolet line emission is
$\sim6.6\times10^{30}$\,\es, about $\sim20$\,\% of the sum of the
optical disk luminosity and the X-ray luminosity
\citep{gaensickeetal97-1}.  Including the ultraviolet disk emission in
the energy balance of the system does, therefore, not noticeably
change the conclusion of \citet{gaensickeetal97-1} that the accretion
rate in EK\,TrA is a factor $\sim5$ lower than in the prototypical
SU\,UMa dwarf nova VW\,Hyi.

It is interesting to compare the short-term ultraviolet variability of
EK\,TrA with that of a long-period dwarf nova. 
\citet{hoardetal97-1} analysed fast HST/FOS spectroscopy of IP\,Peg
and found a strong contribution of the continuum to the ultraviolet
flickering. This indicates that for the higher accretion rates
prevailing above the period gap the disk/corona is a significant
source of ultraviolet continuum emission. Indeed, in IP\,Peg the white
dwarf emission is not detected in the
ultraviolet. \citet{hoardetal97-1} could also show that in IP\,Peg the
flickering in the continuum, medium-excitation, and high-excitation
lines is uncorrelated, suggesting a rather complex multi-component
spectrum from a structured emission region.

\begin{table}
\begin{minipage}{84mm}
\begin{flushleft}
\caption{\label{t-rotation_rates}
White dwarf rotation rates measured from Doppler-broadened
absorption lines, and effective temperatures measured in deep quiescence.}
\begin{tabular}{rrrrrr}\hline\noalign{\smallskip}
System & Type & \Porb & $v\sin i$ & $\Teff~^{(a)}$ & Ref.\\ 
       &      & [min] & [\kms]    & [K]     &  \\ 
\noalign{\smallskip}\hline\noalign{\smallskip}
WZ\,Sge & DN/WZ &  81.6 &     1200  & 14\,900 & 1\\
EK\,TrA & DN/SU &  90.5 &      200  & 18\,800 & 2\\
VW\,Hyi & DN/SU & 106.9 &      400  & 19\,000 & 3,4\\
U\,Gem  & DN/UG & 254.7 & $\le100$  & 32\,000 & 5,6\\
RX\,And & DN/UG & 302.2 &      150  & 35\,000 & 7\\
\noalign{\smallskip}\hline\noalign{\smallskip}
\multicolumn{6}{p{8cm}}{
(1) \citealt{chengetal97-1}; (2) this paper;
(3) \citealt{sionetal95-1}; (4) \citealt{gaensicke+beuermann96-1}; (5) 
\citealt{sionetal94-1}; (6) \citealt{longetal95-1}; (7) \citealt{sionetal01-1}.}

\end{tabular}
\end{flushleft}
\end{minipage}
\end{table}

\section{Discussion \& Conclusions}
EK\,TrA was selected as a target for our HST program as it appears to
be very similar to the well-studied SU\,UMa dwarf nova VW\,Hyi
\citep{gaensickeetal97-1}. Our analysis of the STIS data confirms this
suggestion: we find a white dwarf temperature and rotation rate that
are very close to the values derived for VW\,Hyi.

EK\,TrA is only the fifth CV white dwarf whose rotation rate could be
accurately measured from the Doppler-broadened metal lines in
high-resolution ultraviolet spectra, and it is only the second typical
SU\,UMa-type system (Table\,\ref{t-rotation_rates}). The other four
stars include the ultra-short period large-outburst amplitude dwarf
nova WZ\,Sge, the prototypical SU\,UMa dwarf nova VW\,Hyi, and the two
long-period U\,Gem-type dwarf novae U\,Gem and RX\,And. If we consider
the membership to one of these three groups as a measure of the
evolutionary stage of a system, with the U\,Gem type stars above the
orbital period gap being the youngest systems, the SU\,UMa stars below
the gap being significantly older, and the WZ\,Sge stars being the
oldest -- possibly containing already degenerate donors and evolving
to longer orbital periods -- then the presently known white dwarf
rotation rates are not in disagreement with an increase of
the rotation rate with increasing age. Such a trend is indeed
expected, as the angular momentum of accreted matter spins the white
dwarf up \citep{kingetal91-1}, even though the long-term angular
momentum evolution of accreting white dwarfs --~taking into account
nova explosions~-- is not yet well understood
\citep{livio+pringle98-1}.

\acknowledgements{We thank the CTIO Director's Office for the allocation of
discretionary time used for this project.  CTIO is operated by AURA,
Inc., under cooperative agreement with the United States National
Science Foundation. BTG was supported by the DLR under grant
50\,OR\,99\,03\,6. PS, EMS, and SBH acknowledge partial support of this research
from HST grant GO-08103.03-97A.}

\bibliographystyle{apj}

\end{document}